# RADIATION DAMAGE AND LONG-TERM AGING IN GAS DETECTORS[1]

MAXIM TITOV

*Institute of Physics, Albert-Ludwigs University of Freiburg, Hermann-Herder Str. 3, Freiburg, D79104, Germany and*

*Institute of Theoretical and Experimental Physics (ITEP), B. Cheremushkinskaya, 25, Moscow, 117259, Russia*

Aging phenomena constitute one of the most complex and serious potential problems which could limit, or severely impair, the use of gaseous detectors in unprecedented harsh radiation environments. Long-term operation in high-intensity experiments of the LHC-era not only demands extraordinary radiation hardness of construction materials and gas mixtures but also very specific and appropriate assembly procedures and quality checks during detector construction and testing. Recent experimental data from hadron beams is discussed. It is shown that the initial stage of radiation tests, usually performed under isolated laboratory conditions, may not offer the full information needed to extrapolate to the long-term performance of real and full-size detectors at high energy physics facilities. Major factors, closely related to the capability of operating at large localized ionization densities, and which could lead to operation instabilities and subsequent aging phenomena in gaseous detectors, are summarized. Finally, an overview of aging experience with state-of-the-art gas detectors in experiments with low- and high-intensity radiation environments is given with a goal of providing a set of rules, along with some caveat, for the construction and operation of gaseous detectors in high luminosity experiments.

## 1. Introduction

Recent high energy physics experiments require development of radiation hard gaseous detectors able to withstand hadron fluences up to $10^{15}$ -$10^{16}$ $cm^{-2}$ at the Large Hadron Collider (LHC) and at future very high luminosity colliders. Full functionality of large-area systems over the lifetime of an experiment in a harsh radiation environment is of a prime concern to the involved experimenters. This has resulted in a large research effort aimed at better understanding and improving the radiation hardness of existing proportional chambers and straw-type detectors, which must be able to tolerate considerable rate of heavily ionizing particles and dramatic increase in the charge (up to 1 C/cm/wire per year), accumulated on the sensing electrodes in the future high rate experiments. A new generation of more powerful Micro-Pattern Gaseous Detectors (MPGD), such as Gas Electron Multipliers (GEM), Micromegas and related detector types, which rapidly evolved with the introduction of microelectronic fabrication techniques, has very promising performances with increased reliability under harsh radiation conditions. The diversified research of their aging properties revealed that they might be even less vulnerable to the radiation-induced performance degradation in the future high rate experiments than the standard silicon microstrip detectors.

The achievements of past R&D projects in gas detectors are excellently summarized in [1-4], new developments in the radiation damage research have been reviewed at the 'International workshop on Aging Phenomena in Gaseous Detectors' held at DESY, Hamburg in October, 2001 [5,6]. A great effort has resulted in the optimization and construction of large systems of gas detectors for intense hadron beams, depending on the need for the best spatial resolution (MPGD), best timing resolution for trigger (Resistive Plate Chambers (RPC), MPGD), high-rate tracking and particle identification (straw-type detectors and multi-wire proportional chambers (MWPC)) and single photon detection (CsI-GEM photodetectors). The paper summarizes wide range of aging tests, aimed at increasing reliability and radiation hardness of gas detector technologies in view of their application in future high-energy physics experiments.

## 2. General Characteristics of Aging Phenomena in Wire Chambers

Aging effects in wire chambers, a permanent degradation of operating characteristics under sustained irradiation, has been and still remains a main limitation to the use of gas detectors in high energy physics experiments [7]. The ‘classical aging effects’ are the result of chemical reactions occurring in avalanche plasma near anodes in wire chambers leading to formation of deposits on electrode surfaces [4,8]. During gas avalanches many molecules break up in collisions with electrons, de-excitation of atoms, and UV-absorption processes. Whereas most ionization processes require electron energies greater than 10 eV, the breaking of covalent bonds and formation of free radicals requires only 3-4 eV, and can lead to a higher concentration of free radicals than that of ions in gaseous discharges. (Free radicals are unionized atomic or molecular species with one or more unsatisfied valence bonds). Consequently, free-radical polymerization is regarded as the dominating mechanism of wire chamber aging. Since free radicals are chemically very active they will either recombine to form the original molecules of other volatile species, or may start to form new cross-linked molecular structures of increasing molecular weight. In general, the rate of polymer formation depends upon many microscopic variables such as cross-sections of electron and photon processes and their energy distributions in gas avalanches, molecular dissociation energies as well as densities of electrons, ions and free radicals. The ‘classical aging effects’ lead to the formation of deposits, conductive or insulating, on the electrode surfaces and manifest themselves as a decrease of the gas gain due to the modification of the electric field, excessive currents, sparking and self-sustained discharges. The radiation-induced degradation in wire chambers is sensitive to the nature and purity of the gas mixture, different additives and trace contaminants, materials used in contact with the gas, materials of electrodes and configuration of electric field. In the following, several examples of ‘classical aging effects’, observed in large systems at ‘low rates’ (collected charge ~ mC/cm/wire per year), are discussed from the vast majority of aging data.

---

[1] Invited talk, will be published in the Proceedings of the 42nd Workshop of the INFN ELOISATRON Project `Innovative Detectors for Supercolliders', September 28 – October 4 (2003), Erice, Italy.

- There are a lot of experiments that clearly indicate premature aging in Ar/$CH_4$ mixtures [2,4,5,9]. This observation indicates that $CH_4$ itself polymerizes in the avalanche plasma due to the hydrogen deficiency of radicals and their ability to make bonds with hydrocarbon molecules, and similarly for all hydrocarbon gases. The aging rate in Ar/hydrocarbon gases can be reduced by addition of oxygen-containing molecules, allowing large systems to operate at low intensity without dramatic loss of performance [11-13]. However, Ar/hydrocarbon gases are not trustworthy for long-term, high rate experiments.

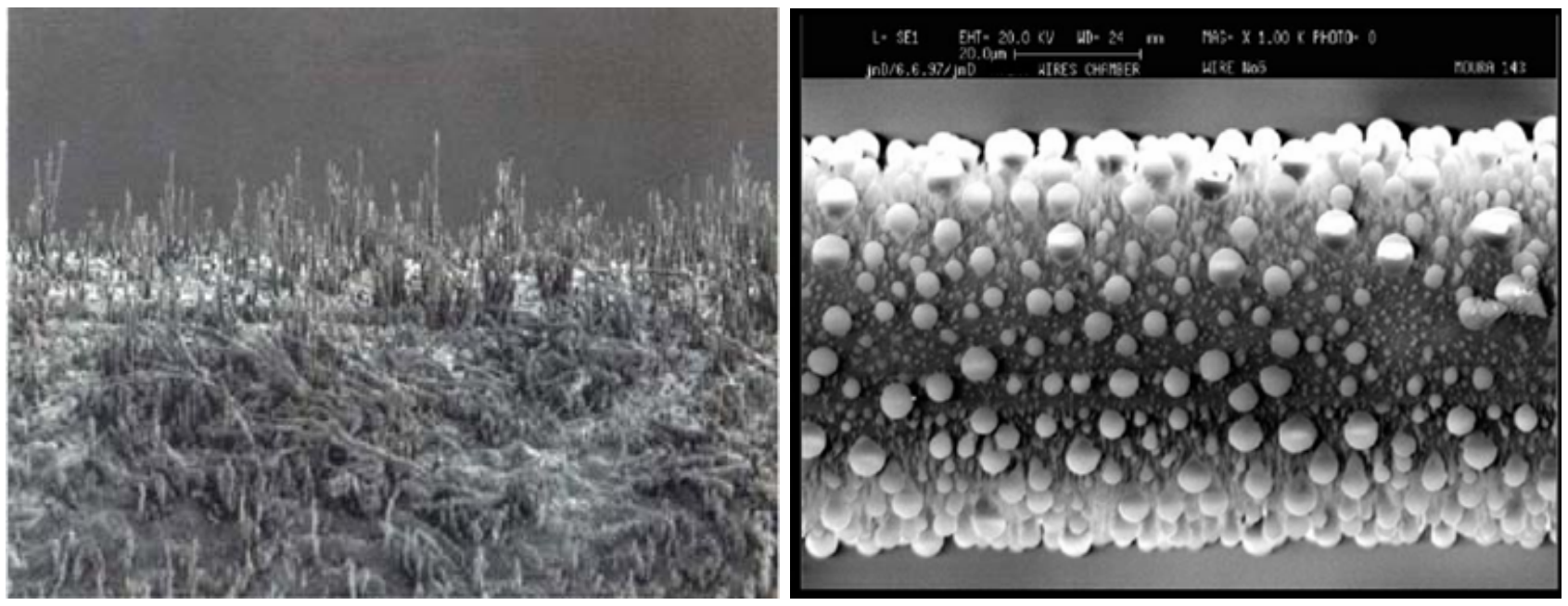

Figure 1. Examples of Si-deposits on the anode wires. (Left) picture from [13]. (Right) from [14].

- Silicon has been systematically detected in analysis of many wire deposits (see Fig.1), although in many cases the source of the Si-pollutant has not been clearly identified [2,4]. The Si-compounds are found in many lubricants, adhesives and rubber, encapsulation compounds, silicon-based grease, various oils, G-10, RTV, O-rings, fine dust, gas impurities, polluted gas cylinders, diffusion pumps, standard flow regulators, molecular sieves, and their presence may not necessarily be noted in the manufacturer’s documentation [9]. The most dangerous ones are Si-lubricant traces, which are widely used for the production of gas system components, may be efficiently clean by flushing DME through the gas system. If the amount of Si-pollutant is not very large, the equipment (e.g. flow-regulator parts) can be cleaned in an ultrasonic bath with isopropyl alcohol [15]. Due to the high specific polymerization rate of Si in gas detectors, a strict validation process for all assembly components has to be foreseen [16].

- The Malter effect [17] is due to the microscopic insulating layer deposited on a cathode from the quencher dissociation products and/or pollutant molecules. Some of the metal oxide coatings, adsorbed layers on the cathode or cathode material itself may not be initially conducting enough and may inhibit the neutralization of positive ions arriving from the avalanche, which generate a strong electric field across dielectric film and cause electron field-emission from the cathode [8]. A positive feedback between the electron emission at the cathode and anode amplification will lead to the appearance of dark current, increased rate of noise pulses and finally may result in the exponential current growth (classical Malter breakdown) [18]. The addition of water was found to prevent Malter discharges in classical wire chambers, since water increases the conductivity of partially damaged electrodes; in one test Malter effect was cured by the addition of $O_2$ to damaged chamber and running at large current [19].

## 2. Aging phenomena in the high-rate LHC era

A dramatic increase of the radiation intensity encountered by gaseous detectors (collected charge ~ C/cm/wire per year) at the high-rate experiments of the LHC era has demanded a concerted effort towards finding appropriate gas mixtures, adequate assembly materials and procedures and strict quality control checks as a part of fighting against aging. Very often, problems encountered with first prototype detectors for the LHC environments have been due to the casual selection of chamber designs, gas mixtures, materials and gas system components, which successfully worked with certain gas gain at ‘low rates’ [4], but rapidly failed in harsh radiation environment [5]. For example, many non-metallic ‘good materials’ successfully used in the large systems of the LEP era might nevertheless outgass at a small level, thus causing fast aging under high rates [3,9]. A large amount of outgassing data have been accumulated in the framework of project RD-10 (A study to improve the radiation hardness of gaseous detectors for use at very high luminosity), which afterwards was merged with the more specific research RD-28 on MSGC (Development of Micro-Strip Gas Chambers for radiation detection and tracking at high rates) [20]. A specific material is either adequate or not for a particular detector type and operating conditions – one has to do tests matching the specific requirements of the experiment. In the following, a summary of recent research, focused on the understanding of detector behavior after exposure to the very high radiation levels anticipated in the LHC-like conditions, will be described.

### 2.1. *Gases of possible interest for the high-rate hadronic environments*

Future high energy and luminosity colliders pose a new challenge for gas mixtures, raising the requirement to their radiation hardness up to ~10 C/cm/wire for wire-type and ~20 $mC/mm^2$ for micro-pattern detectors. While a huge variety of gases have been successfully used for the `standard radiation level’ detectors, only a limited choice of gases is available for the new high-rate experiments of the LHC era. Among the conventional gas mixtures only $Ar(Xe)/CO_2$ and $Ar(Xe)/CO_2/O_2$ are demonstrated to tolerate such doses. Unfortunately, $Ar(Xe)/CO_2$ are quite transparent for photons. The addition of 2-3% $O_2$ enlarges the operational safety margin of the mixture (ozone is a strong absorber of UV-photons), while reducing the signal due to the electron attachment [16].

About 20 years ago, $CF_4$ was proposed as the most attracting candidate for high-rate environments [21,22]. This is primarily due to the high-drift velocity, high primary ionization, low electron diffusion and resistance to aging [23-25]. However, $CF_4$ molecule has a small quenching cross-section of metastable Ar-states [26] and excited $CF_4$ molecules copiously emit photons from the far UV (peak around ~170 nm – 7.3 eV) to the visible light, and have about 25% of Ar scintillation efficiency in the UV [27,28]. Being transparent to its own UV photons, $CF_4$ presents serious gain limitation in wire chambers due to the photo-electric emission from the cathode. On the contrary, GEMs operate well in $Ar/CF_4$ because the avalanche confinement within the GEM holes strongly suppresses photon-mediated secondary processes. In addition, $Ar(Xe)/CF_4$ mixtures have rather poor energy and spatial resolution due to the electron attachment processes in $CF_4$, which occurs mainly in the 6 to 8 eV range [29,30]. The production of long-lived and highly electronegative radicals, observed under strong irradiation in $CF_4$-based mixtures, can significantly degrade performance of large detectors with insufficient gas flow or operated in a closed-loop gas system [31-35]. The advantages of the enhanced drift velocity of $CF_4$ for high rate applications have been realized by the addition of one of the common quenchers (e.g. $CH_4$,$CO_2$) to $CF_4$ or to Ar/$CF_4$. This can also 'cool' electrons to the extent that attachment does not occur [23].

There has been a strong interest in the use of $CF_4$ for high-rate applications in future high luminosity colliders because it was believed that most of $CF_4$-based mixtures can prevent 'classical aging effects'. Since long time $CF_4$ is used in plasma processing as an excellent etching gas for Si and Si-containing materials, which react with fluorine to form volatile Si-F based compounds [36]. However, $CF_4$ plasma is not a good etching gas for organic polymers that do not contain Si in the backbone chain and contain lots of H. Actually, the addition of hydrogenated species shifts the chemistry of $CF_4$ plasmas towards polymerization (Ar/$CF_4$/$CH_4$ and $C_2H_2F_4$ gases would form polymers, at least at low pressure) [37,38], while addition of oxygenated species shifts the chemistry of $CF_4$ plasmas towards etching [39]. While plasma characteristics (50 mtorr, rf, deposition rate ~50 nm/min) differ from those of gas detectors (atm, dc, ~50 nm/year), it is likely that some useful hints of plasma chemistry could be applicable to the field of gaseous detectors (energy range of electrons is similar) [1,2]. In wire chambers, within the broad spectrum of gases there is no mixture without $CF_4$ that is able to tolerate doses ~ 20 C/cm/wire [15]. However, the presence of a $CF_4$ component in the gas limits the choice of construction materials for detectors and gas systems. Decomposition of $CF_4$ molecules in the avalanche process contaminates the gas with fluorine radicals, which react violently with exposed electrode surfaces to form metal fluorides or cause etching phenomena. Non-gold anode wires are unacceptable for use in $CF_4$-based mixtures due to the formation of metal fluorides [40]. However, the use of gold-plated electrodes does not necessarily ensure good aging properties in $CF_4$, and many results show contradictory experience with $CF_4$-based operation, where both polymerization and material etching phenomena take place.

### 2.2. *Aging in Ar/$CF_4$/$CH_4$, $CF_4$/$CH_4$ and $C_2H_2F_4$ mixtures*

Several laboratory tests in the beginning of the 90's have demonstrated excellent aging properties, up to 10 C/cm/wire, of $CF_4$/i$C_4H_{10}$ (80:20) avalanches, which also had an ability to etch silicon-based and hydrocarbon deposits, from previously etched Au/W-wires [40-42]. Based on these studies, the dominant role of etching processes with increasing $CF_4$ concentration was assumed for $CF_4$/hydrocarbon mixtures. However, using the same laboratory setup for accelerated tests (current density ~ 1.2μA/cm), a lack of apparent aging has been observed in $CF_4$/i$C_4H_{10}$ mixtures for $CF_4$ concentrations (30-85% and ~100%), while heavy carbonaceous deposits were observed for other $CF_4$ concentrations on Au/W-wires [40]. The existence of an aging region for $CF_4$ concentrations between ~85% and <100% was unexpected. The authors proposed a model to explain two regions of polymerization and etching in $CF_4$/i$C_4H_{10}$, based on the relative proportions of $CF_4$ and i$C_4H_{10}$ in the mixture. However, a quick formation of anode deposits in the 'aging resistant' $CF_4$/i$C_4H_{10}$ (80:20) gas was observed in the high rate environment [43]. Several other studies under harsh radiation environment have demonstrated that $CF_4$/hydrocarbon gases represent the most extreme case of etching competiting with polymer formation. While prototype honeycomb drift chambers have been proven to be immune to very high X-ray doses up to 5 C/cm/wire, severe anode and cathode aging effects were found in Ar/$CF_4$/$CH_4$ (74:20:6) and $CF_4$/$CH_4$ (80:20) mixtures in the high-rate HERA-B environment (secondaries from interactions of 920 GeV protons with target nucleus) [34]. An extensive R&D program, carried out at 10 different radiation facilities to resemble HERA-B conditions, traced aging problems to a combination of several effects; a solution which uses gold-coated cathode foils and Ar/$CF_4$/$CO_2$ mixture was found to survive in the high-rate hadronic environment. The aging performance of HERA-B muon proportional chambers with Ar/$CF_4$/$CH_4$ and $CF_4$/$CH_4$ mixtures has shown that the aging rate depends on the type of irradiation, high voltage and area of irradiation. Fig.2 shows SEM micrographs of typical wire deposits after irradiation in Ar/$CF_4$/$CH_4$ (67:30:3) [35]. A dramatic degradation of the kapton XC foils used as a straw material in the LHC-b tracker was observed under the glow discharge in Ar/$CF_4$/$CH_4$ [44]. These results indicate that $CF_4$/hydrocarbon mixtures might be inclined to growth of polymers at high ionization densities, due to simultaneous polymerization of $CF_x$ and $CH_x$ radicals. Apparently, $C_2H_2F_4$-containing gases used in RPCs are also prone to polymerization; furthermore the addition of water to this gas will greatly enhance the formation of HF, which is a good etching agent of glass in laboratory. In a view of very small safety margins, long-term operation of large systems with $CF_4$/hydrocarbon-based gases in the high radiation density environment of future experiments seems to be of considerable risk.

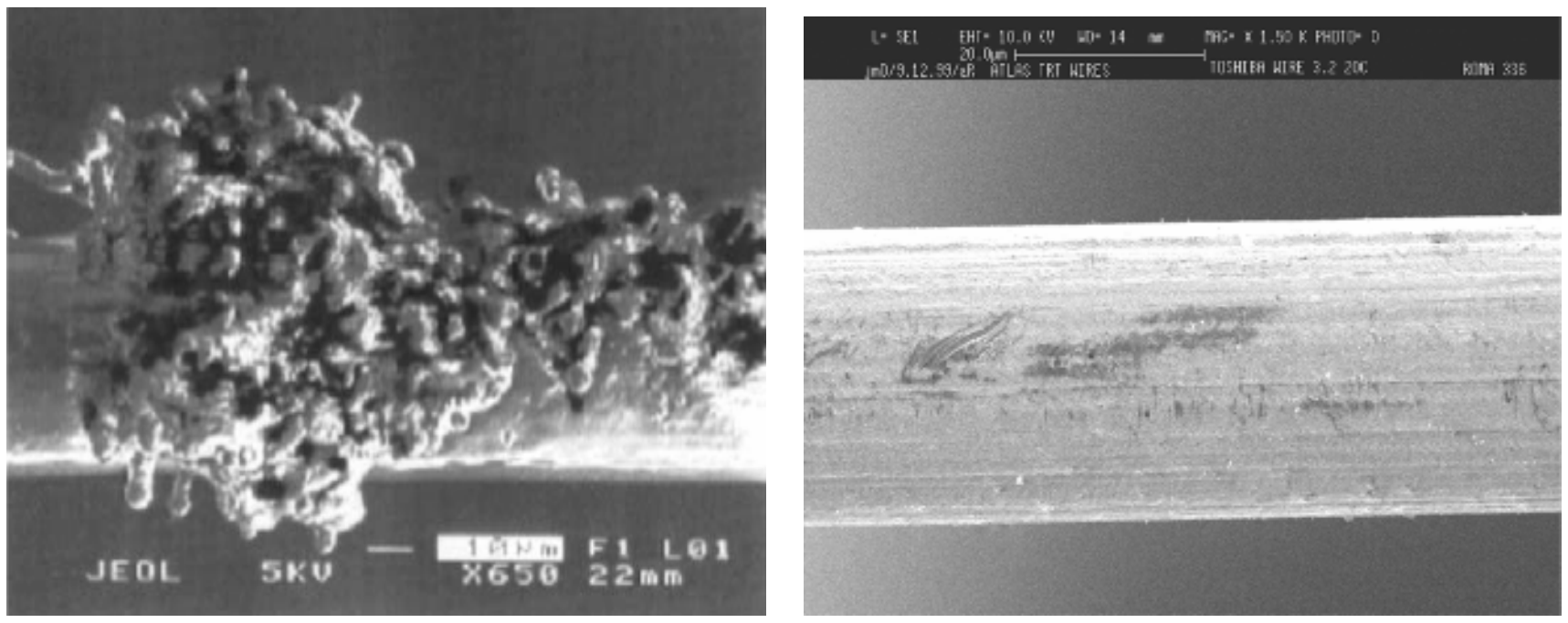

Figure 2. (Left) EDX analysis of the anode wire after 0.07 C/cm/wire in $Ar/CF_4/CH_4$ (67:30:3) mixture shows deposits, consisting only of C and F elements (H is not detectable)[35]. (Right) Toshiba wire surface (Au thickness 0.5μm) after 20 C/cm/wire in $Xe/CF_4/CO_2$ (70:20:10) [15].

### 2.3. *Aging in $Ar/CF_4$ and $Ar/CF_4/CO_2$ mixtures*

Both $Ar/CF_4$ and $Ar(Xe)/CF_4/CO_2$ seem to be perfect for non-polymerizing gases; under optimal operating conditions, no observable drop in gain due to polymerization of gas mixture components has been found in many tests (see Table 1). Moreover, a small addition of $CF_4$ to $Ar/CO_2$ in the high luminosity LHC environment is very efficient to clean up Si-aged wires [16]. The resulting balance between Si-polymerization and $CF_4$ etching processes was found to be very sensitive to the Si-source intensity and ionization density [15]. Recent investigations in $Ar(Xe)/CF_4/CO_2$ mixtures revealed 'new aging phenomena' – the etching of the gold-plating of anode wires in straws and rupture of wires in honeycomb drift tubes [15,34,45]. Dedicated studies for the ATLAS TRT straws demonstrated that the main components responsible for the gold wire damage, and formation of WO deposits, are chemically reactive products of $CF_4$ disintegration, in connection with water (HF acid) [15]. The magnitude of the damage processes was found to depend on the wire manufacturing technology and gold plating thickness. No gold damage effects have been observed, for Toshiba 30 μm Au/W-wire and $H_2O$ < 0.1%, up to 20 C/cm (see Fig.2). Moreover, destruction processes of wire gold-plating were usually observed at collected charges larger than 0.5 C/cm/wire accumulated at extreme current densities (~1-5 μA/cm), compared to the LHC environment (<100nA/cm). Initial imperfections in the gold crystalline structure might stimulate local micro-discharges and sparking under very high dose rates, which can greatly enhance the magnitude of $CF_4$ plasma etching processes compared to the typical avalanche conditions and moderate charge accumulation rates, and play a dominant role in further destruction of gold-plating. No significant gold damage effects have been observed for several wire types at current densities <0.5 μA/cm (see Table 1).

While no anode deposits, due to the polymerization of gas mixture components, have been observed on Au/W-wires after exposure in $Ar/CF_4/CO_2$, a bulk of data indicate the changes of consistency and appearance of fluorine-containing (often accompanied by Si) films on non-gold plated cathodes [14,35,40,46,47]. Since the resultant insulating layers may lead to increased dark current, spurious noise counts and finally trigger the self-sustained Malter discharge, the use of the more robust electrode materials to $CF_4$ is recommended (e.g. gold-plated electrodes or cathode straw materials) [15,52]. Another very dangerous drawback of the $CF_4$ avalanches is creation of harmful fluorine radicals and HF acid, which were found to attack and strongly etch surfaces exposed to gas, such as Al and Cu coatings in straw, diamond coated MSGC, GEM foils, and heavily destroy all glass materials (rotameters, FR4, etc.) [16,41,48-51]. For instance, the glass wire joints (containing Si) inside long straws were etched to the point of breakage in $Xe/CF_4/CO_2$; a new $Xe/CO_2/O_2$ mixture has been selected to ensure reliable operation in the ATLAS TRT.

Nevertheless, short periodic 'cleaning runs' with $Ar/CF_4/CO_2$ (70:3:27) in a flushing mode are foreseen, if needed, to clean up Si-aged straw wires [16]. In general, it seems that the $CF_4$ etching rate is roughly proportional to the $CF_4$ concentration and to the current density.

Table 1. Aging results with wire detector prototypes for the use at high energy physics facilities.

| Experiment, Detector, Reference | Gas Mixture | Gain reduction/ etching | Cathode aging | Charge C/cm | Gas gain(G); Current density(I); Rate (R); Irradiated area (S) |
|---|---|---|---|---|---|
| **HERA-B OTR [34]** | **$Ar/CF_4/CO_2$ (65:30:5)** | **No/Au damage** | **No** | **0.6** | **$G\sim3*10^4$; $I\sim0.4$-$0.9\mu A/cm$; 100MeV $\alpha$-beam; $S\sim50cm^2$** |
| **HERA-B MUON [35]** | **$Ar/CF_4/CO_2$ (65:30:5)** | **No** | **No/F-film on Al cath.** | **0.7** | **$G\sim10^4$-$10^5$; $I\sim0.3\mu A/cm$; $R<10^5$ Hz/cm$^2$; $S\sim1200cm^2$** |
| **HERA-B MUON[35]** | **$Ar/CF_4$ (70:30)** | **No** | **No** | **0.07** | **$G\sim10^5$; $I\sim0.15\mu A/cm$; $R\sim10^4$ Hz/cm$^2$; $S\sim150cm^2$** |
| **ATLAS TRT [15,48]** | **$Xe/CF_4/CO_2$ (70:20:10)** | **No** | **No** | **20** | **$G\sim 3*10^4$; $I\sim0.7\ \mu A/cm$; $S\sim24\ cm^2$** |
| **Straw R&D[45]** | **$Xe/CF_4/CO_2$ (70:20:10)** | **No/Au cracks** | **No** | **9** | **$G\sim3*10^4$; $I\sim1.7\mu A/cm$; $R\sim2*10^6$ Hz/cm$^2$; $S\sim1cm^2$** |
| **Straw R&D[45]** | **$Ar/CF_4/CO_2$ (60:10:30)** | **No/Au cracks** | **No** | **1.5** | **$G\sim10^5$; $I\sim1.7\mu A/cm$; $R\sim10^6$ Hz/cm$^2$; $S\sim1cm^2$** |
| **CMS MUON [14]** | **$Ar/CF_4/CO_2$ (40:10:50)** | **No/ Au cracks** | **Si-F film on Cu cath.** | **13.4** | **$G\sim6*10^4$; $I\sim2\mu A/cm$; $R\sim2*10^6$ Hz/cm$^2$; $S\sim1cm^2$** |
| **CMS MUON [46]** | **$Ar/CF_4/CO_2$ (40:10:50)** | **No** | **Si-F film on Cu cath.** | **0.4** | **$G\sim10^5$; $I<0.05\ \mu A/cm$; $R\sim2*10^4$ Hz/cm$^2$;$S\sim21000cm^2$** |
| **LHC-B MUON [52]** | **$Ar/CF_4/CO_2$ (40:10:50)** | **No** | **Etching of FR4 bars** | **0.25** | **$G\sim10^5$; $I<0.03\ \mu A/cm$; $R\sim3*10^4$ Hz/cm$^2$; $S\sim1500cm^2$** |
| **COMPASS Straws [53]** | **$Ar/CF_4/CO_2$ (74:20:6)** | **No/ Si-traces** | **No** | **1.1** | **$G\sim4*10^4$; $I\sim4\ \mu A/cm$; $R\sim2*10^8$ Hz/cm$^2$; $S\sim3\ cm^2$** |
| **HERMES FD [47]** | **$Ar/CF_4/CO_2$ (90:5:5)** | **No** | **Al etching/ Cl deposits** | **9** | **$G\sim5*10^5$; $I\sim1\ \mu A/cm$; $R\sim10^5$ Hz/cm$^2$; $S\sim7\ cm^2$** |
| **ATLAS TRT[16]** | **$Xe/CO_2/O_2$ (70:27:3)** | **No** | **No** | **11** | **$G\sim 3*10^4$; $I\sim1$-$3\ \mu A/cm$; $S\sim1\ cm^2$** |
| **ATLAS TRT[16]** | **$Xe/CO_2/O_2$ (70:27:3)** | **No/Si-deposits** | **No** | **0.3** | **$G\sim3*10^4$; $I\sim0.1\ \mu A/cm$; $S\sim1\ cm^2$** |
| **ATLAS MDT[55]** | **$Ar/CO_2$ (90:10)** | **No** | **No** | **0.7** | **$I\sim\ 0.1\mu A/cm$; $R\sim4*10^2$ Hz/cm$^2$;$S\sim7500cm^2$** |
| **ATLAS MDT[56]** | **$Ar/CO_2$ (93:7)** | **No** | **No** | **>1.5** | **$G\sim2*10^4$; $I\sim0.05$-$0.2\mu A/cm$; $R\sim8*10^2$-$10^4$Hz/cm$^2$;$S\sim90cm^2$** |
| **ATLAS MDT[57]** | **$Ar/CO_2$ (93:7)** | **Yes/Si-deposits** | **No** | **0.2** | **$G\sim9*10^4$; $I\sim\ 0.02\mu A/cm$; $R\sim5*10^2$ Hz/cm$^2$;$S\sim15000cm^2$** |
| **CMS MUON[14]** | **$Ar/CO_2$ (30:70)** | **Yes/ Si-deposits** | **No** | **0.76** | **$G\sim6*10^4$; $I\sim\ 2\ \mu A/cm$; $R\sim2*10^6$ Hz/cm$^2$; $S\sim1\ cm^2$** |

Finally, the use of very expensive $Xe/CF_4$-based gases forces large detectors to use recirculation systems where special purification elements to remove long-lived fluorine radicals have to be foreseen. For the 'cheaper' $Ar/CF_4$ gases, accumulation of fluorine molecules in the closed-loop system can be avoided by replacing part of the mixture coming out from detector by fresh gas, or operation in a flushing mode, at the expense of extra costs, can be considered. The very high aggressiveness of $CF_4$ dissociative products and the dynamic modification of the gas composition require more studies to evaluate the impact of these effects on the long-term performance and stability of large-area gaseous detectors. However, $CF_4$ is very effective in blocking the extremely negative influence of Si-pollutants; the apparent immunity of $CF_4$-containing mixtures to the classical aging effects might play a unique role for the improved radiation hardness of gas detectors in severe radiation LHC environment.

### 2.4. *Aging in Ar(Xe)/$CO_2$ and Ar(Xe)/$CO_2$/$O_2$ mixtures*

Both Ar(Xe)/$CO_2$ and Ar(Xe)/$CO_2$/$O_2$ could be absolutely radiation resistant under clean conditions in all types of gaseous detectors; there is no well-established mechanism which could lead to formation of anode deposits in these mixtures [1,2,54-56,107]. However, the aging performance in Ar(Xe)/$CO_2$ and Ar(Xe)/$CO_2$/$O_2$ is sensitive to hydrocarbon contamination and even to minor traces of silicon pollutants in the detector (e.g. straw material may contain a few atomic Si mono-layers) and in the gas-system components (Si-based lubricant residues) [14-16,57]. In case of a weak external silicon source in the gaseous detector, several studies revealed appearance of aging effects at the beginning (in terms of gas flow) of the irradiated area (all Si-molecules are quickly eaten up by the avalanche process and produce deposits) and sometimes even right before the irradiated area (due to the upstream diffusion of active species at low gas flows). In case of much larger pollution by Si-lubricants, deposits have a tendency to propagate along the whole detector.

In recent tests, silicon deposits were also observed in $Xe/CO_2/O_2$ in the gas system, which was successfully operated during several years with the $CF_4$-containing mixture without any aging effects [16]. This would indicate that even specially certified and cleaned gas systems for $CF_4$ operation might have small traces of silicon contamination. Therefore, extraordinary validation procedures will be required to reach the necessary level of purity for all components used in the detectors and closed-loop recirculation gas systems in $Ar(Xe)/CO_2$; the final prove that $CF_4$-free gases can be used for large-scale, high-rate, long-term applications still has to come.

## 3. Aging in Micro-pattern Gas Detectors (MPGD)

Future high-luminosity experiments have prompted a series of inventions of novel high-rate gaseous detectors: MSGC, GEM, MICROMEGAS and many others [58]. The MSGC, a novel concept invented in 1988 by Oed [59], was the first of the microstructure gas detectors. Consisting of a set of tiny metal strips laid on a thin insulating substrate, and alternatively connected as anodes and cathodes, the MSGC relies for its operation on the same processes of avalanche multiplication as multi-wire devices [60]. Despite the promising performance (spatial resolution ~40μm, high rate capability $<10^6$ Hz mm$^{-2}$), MSGC has entered a new dimension of sensitivity to aging, compared to MWPC, due to the filigree nature of MSGC structures and catalytic effects on the MSGC substrate. There are several major processes, particularly at high rates, leading to MSGC operating instabilities: charge-up of substrates, surface deposition of polymers, and destructive microdischarges [3,60,61]. The influence of glass conductivity has been verified for MSGCs: the use of ionic conductive glass DESAG D-263 as a substrate results in rate-dependent modification of gain due to the radiation induced variation of surface resistivity. This effect can be explained as electrical field distortions caused by accumulation of charges on the insulating surface between strips and/or slow migration of ions (the majority charge carriers) inside the glass bulk. Using MSGCs manufactured on electron-conducting SCHOTT S-8900 or diamond-coated D-263 glass, excellent high-rate performance and long-term stability have been demonstrated. Early aging studies of MSGCs indicated that they are much more susceptible to aging than wire chambers, possibly due to a small effective area used for charge multiplication and thus higher energy density in the avalanche plasma [3]. Micro-strip chambers have been successfully operated with a large variety of gases; to prevent fast aging at high rates, convincing evidence suggests to avoid organic quenchers in the gas. Under optimal laboratory conditions, a lifetime of the MSGC equivalent to more than 10 years of operation in the LHC has been achieved (see Table 2). However, although MSGC's survive very strong particle fluxes of X-rays or electrons, they can be destroyed within a few hours in beams of pions or protons due to multiple anode-cathode discharges (sparks). The problem of microdischarges, induced by heavily ionizing particles and destroying the fragile MSGC electrode structure, turned out to be a major limitation to all single-stage micro-pattern detectors in hadronic beams [51,62].

Sharing the amplification process between two cascaded devices, a GEM followed by MSGC, permits, for a given gain, to operate both elements below the onset of induced discharges in a high rate hadronic environment. Introduced by Sauli in 1996, a GEM consists of a set of holes, typically 50-70 μm in diameter, chemically etched through a 50 μm thick copper-kapton-copper polymer foil [63]. Application of a suitable voltage difference between the metal layers of the GEM produces a strong electric field in the holes (~50-100 kV/cm), where the gas amplification happens. No aging effects and no damage of electrode strips were observed for MSGC-GEM in Ar/DME in high-rate X-ray tests, irradiating 'small' surface areas.

However, the use of Ar/DME, originally foreseen as a counting gas in most MSGC and MSGC+GEM developments, lead to visible deposits on electrodes under X-rays, if the size of irradiated area is large enough (~$10^4$ mm), while identical chambers with Ar/$CO_2$ showed no aging [51]. The large MSGC+GEM detectors have been successfully used in the DIRAC [64] and in the HERA-B experiments [51].

Table 2. Summary of aging experience with Micro-Pattern Gas Detectors.

| Detector type | Mixture | Gain reduction ΔG/G | Charge mC/mm$^2$ | Current density, nA/mm$^2$ | Irradiated area; Irradiation rate | Irrad. source |
|---|---|---|---|---|---|---|
| **MSGC [65,104]** | **Ar/DME (90:10)** | **No** | **200** | **10** | **3 mm$^2$; $10^5$ Hz/mm$^2$** | **6.4 keV X-rays;** |
| **MSGC [66]** | **Ar/DME (50:50)** | **No** | **40** | **63** | **0.3*0.8 mm$^2$; 8*$10^5$ Hz/mm$^2$** | **5.4 keV X-rays** |
| **MSGC [66]** | **Ar/DME (50:50)** | **Anode deposit** | **50** | **9** | **0.3*0.8 mm$^2$; 2*$10^4$ Hz/mm$^2$** | **5.4 keV X-rays** |
| **MSGC+GEM [66]** | **Ar/DME (90:10)** | **No** | **70** | **63** | **0.3*0.8 mm$^2$; 2*$10^6$ Hz/mm$^2$** | **5.4 keV X-rays** |
| **MSGC+GEM [51]** | **Ar/DME (50:50)** | **No** | **5** | **2** | **~ 113 mm$^2$; 2*$10^4$ Hz/mm$^2$** | **8 keV X-rays** |
| **MSGC+GEM [51]** | **Ar/DME (50:50)** | **Yes** | **0.5** | **1** | **~ 900 mm$^2$; $10^4$ Hz/mm$^2$** | **8 keV X-rays** |
| **MSGC+GEM [51]** | **Ar/$CO_2$ (70:30)** | **No** | **10** | **2** | **350 mm$^2$ 2*$10^4$ Hz/mm$^2$** | **8 keV X-rays** |
| **Double GEM [67]** | **Ar/$CO_2$ (70:30)** | **Slight gain loss** | **25** | **6** | **2 mm$^2$; 7*$10^4$ Hz/mm$^2$** | **5.4 keV X-rays** |
| **Triple GEM [68]** | **Ar/$CO_2$ (70:30)** | **No** | **27** | **13** | **1.5 mm$^2$; 6*$10^4$ Hz/mm$^2$** | **5.4 keV X-rays** |
| **Double GEM [69]** | **Ar/$CO_2$ (70:30)** | **No** | **12** | **4** | **200 mm$^2$; 5*$10^4$ Hz/mm$^2$** | **6 keV X-rays** |
| **Triple GEM [70]** | **Ar/$CO_2$ (70:30)** | **No** | **11** | **10** | **1260 mm$^2$; 2*$10^4$ Hz/mm$^2$** | **8.9 keV X-rays** |
| **Triple GEM [71,72]** | **Ar/$CF_4$/$CO_2$ (60:20:20)** | **< 5 %** | **230** | **270** | **1 mm$^2$; 5*$10^5$ Hz/mm$^2$** | **5.9 keV X-rays** |
| **Triple GEM [72]** | **Ar/$CF_4$/$CO_2$ (45:40:15)** | **< 5 %** | **45** | **160** | **1 mm$^2$; 5*$10^5$ Hz/mm$^2$** | **5.9 keV X-rays** |
| **Triple GEM [73]** | **Ar/$CF_4$/$CO_2$ (45:40:15)** | **Yes/ 55%** | **20** | **20** | **200*240 mm$^2$** | **25 kCi $^{60}$Co** |
| **Triple GEM [72]** | **Ar/$CF_4$/$C_4H_{10}$ (65:28:7)** | **~ 10 %** | **110** | **160** | **1 mm$^2$; 5*$10^5$ Hz/mm$^2$** | **5.9 keV X-rays** |
| **Triple GEM+ CsI [88]** | **$CF_4$ (100)** | **No** | **0.1** | **3** | **100 mm$^2$; $10^7$ Hz/mm$^2$** | **Hg UV Lamp** |
| **Micromegas [74]** | **Ne/ $C_4H_{10}$ (91:9)** | **Yes/ 35 %** | **10** | **50** | **16 mm$^2$; 33Hz spark rate** | **5.3 MeV α's** |
| **Micromegas [75]** | **Ar/ $CF_4$ (95:5)** | **No** | **2** | **10** | **--------** | **8 keV X-rays** |
| **Micromegas [76]** | **$CF_4$/ $C_4H_{10}$ (94:6)** | **No** | **1.6** | **25** | **20 mm$^2$; ~$10^5$ Hz/mm$^2$** | **8 keV X-rays** |
| **Micromegas [76]** | **Ar/ $C_4H_{10}$ (94:6)** | **No** | **18** | **20** | **20 mm$^2$; 2*$10^5$ Hz/mm$^2$** | **8 keV X-rays** |
| **Micromegas+ GEM[77]** | **Ar/$CO_2$ (70:30)** | **No** | **23** | **17** | **3 mm$^2$** | **5.4 keV X-rays** |
| **MSGC+GEM DIRAC [64]** | **Ar/DME (60:40)** | **10 % eff. drop** | **0.01** | **0.05** | **100*100 mm$^2$; 3*$10^4$ Hz/mm$^2$** | **24 GeV protons** |
| **3-GEM [78] COMPASS** | **Ar/$CO_2$ (70:30)** | **No** | **2.3** | **--** | **310*310 mm$^2$; 2*$10^4$ Hz/mm$^2$** | **Compass Beam** |

Another very attractive and promising solution to reduce the probability of hazardous discharges and to minimize aging effects in MPGD is to separate gas amplification (GEM) and readout stage – printed circuit board (PCB) electrodes, which operate at unity gain and serve only as an electron collector. Many studies of GEM radiation hardness were carried out using X-ray beams, generating about 10-20 larger avalanches than MIP. No significant aging effects in Ar/$CO_2$ were observed in a triple GEM study in Purdue after irradiating a few mm$^2$ of GEM up to 27 mC/mm$^2$ [68]. A sign of moderate aging in the double GEM, after 25 mC/mm$^2$, was found in the form of discoloration around the rims of the holes on the lower GEM electrode, adjacent to the charge collection PCB, and may be due to accumulation of hydrocarbon impurities in the gas [67]. Negligible gain variations were also found in small area aging tests for three different $CF_4$-based mixtures after 100 mC/mm$^2$ [72]. However, an evident current drop was observed in the heavily irradiated large scale (20*24cm$^2$) triple GEM in an $Ar/CF_4/CO_2$ (45:40:15) after 20 mC/mm$^2$ (see Table 2). The decrease of the electric field intensity inside the holes, resulting in a smaller gain, was traced both to the $CF_4$ etching of copper and kapton in proximity of GEM hole edges and appearance of CuF metal fluorides in these places, and can be avoided by the sufficient increase in gas flow [73]. In general, radiation damage to the GEM structure was found to be the most significant around the rims of the holes (where the electric fields exceeds the average) and for the last GEM in series (where number and density of avalanche electrons is the largest). No visible damage of PCB surfaces was found in both double and triple GEMs, which suggests that the absence of a high field amplification region eliminates the possibility of forming deposits on the charge-collecting anodes. The superior radiation hardness of multi-GEM devices was demonstrated in large-area aging tests at CERN. Neither loss of gain nor evidence for the formation of deposits was observed in non-polymerizing $Ar/CO_2$ (70:30) mixture up to ~10 mC/mm$^2$ [69,70]; this accumulated charge is sufficient for large-scale and long-term experiments in intense radiation fields. Furthermore, in these studies the GEM detector was found to be less sensitive to aging, caused by impurities and trace contaminants in $Ar/CO_2$, than a single wire proportional counter installed in the same gas line, behind the GEM detector [69]. However, the conditions at high energy physics colliders near the interaction region can not be approximated by a photon beam. The high electric field at the edges of GEM holes, where the copper and Kapton meet, tends to induce micro-discharges under harsh radiation environments, which could slowly damage the copper and kapton. The modification of the GEM electrode geometry, which could be also enhanced under $CF_4$ avalanches, may influence polarization effects in the insulator, resulting in dependence of charging effects from the accumulated charge, in reduction of rate capability and may also limit the plateau length due to the onset of discharges at smaller voltages. However, since multi-GEM detectors allows sharing the gain between several electrodes, they have much higher reliability and can give only one discharge per $10^{12}$ incident protons, which is 4-5 orders of magnitude smaller than for Micromegas.

Introduced in 1996 by Giomataris [79], a micro-mesh gaseous structure (Micromegas) is a two stage parallel plate electrode gas detector, composed of a ~3 mm drift gap (electric field ~ 1 kV/cm) and a 100 μm high amplification region (~50 kV/cm), located between a thin cathode metal grid (micromesh), and the anode readout plane (strips/pads of a conductor, printed on a insulator board). With a single stage amplification detector like Micromegas one can not eliminate completely discharges in the presence of heavily ionizing particles; with the non-polymerizing $Ar/CO_2$ mixture the observed sparking rate was found to be unacceptable. However, Micromegas, used as a standalone device, were demonstrated not to suffer permanent damages even after accumulating $10^7$/mm$^2$ sparks triggered by highly ionizing particles [74]. The high radiation hardness of Micromegas has been demonstrated using an intense X-ray generator even with hydrocarbon mixtures (see Table 2). It may be related to the fact that avalanche electrons are collected by the grounded anode strips, which are analogous to the PCB in GEM devices. Nevertheless, since the aging is known to occur in a parallel plate geometry [80] and may be increased under sparking due to the heavily ionizing background, long-term aging tests of proper functioning of Micromegas under the expected LHC-like environment have to be foreseen. More work is also needed to study corrosive properties of $CF_4$ on the long-term performance of Micromegas and GEMs in the intense hadron beams. A possible way to enhance their aging performance and to reduce the sparking rate of Micromegas is to add another preamplification device like a GEM. One test has shown that a Micromegas+GEM combination exhibits no deterioration in performance in $Ar/CO_2$ up to 23 mC/mm$^2$ [77].

Both Micromegas and GEM devices seems to be adequate for high-rate, long-term experiments opening an important new field of applications of gas filled detectors. The best confirmation of the radiation hardness of the GEM technology for high-rate applications comes from the successful operation of large area triple-GEMs with $Ar/CO_2$ in the COMPASS experiment (accumulated charge ~ 2.3 mC/mm$^2$ since 2002) [78], where the detected particles are mainly muons at a rate $2*10^7$ Hz with a relatively small amount of hadrons $5*10^4$ Hz. Large area Micromegas are also routinely used for particle tracking in the COMPASS beam [81] and in the $3*10^5$ Hz/mm$^2$ kaon beam in the NA48/KABES experiment in $Ne/C_2H_6/CF_4$ (79:11:10).

## 4. Aging in Gaseous Photodetectors

In recent years, there has been considerable work in the field of photon imaging detectors by combining solid CsI-photocathodes (PC) with MWPCs or GEMs to detect single photo-electrons. There are several mechanisms under the usual RICH irradiation conditions, which may change the electronic structure of CsI and cause a decay of quantum efficiency: exposure to air ($H_20$), photon impact, ion bombardment and creation of polymerizing contaminants in the avalanche process [82-85]. Non-aging gas mixtures are of primary importance for the development of gas-photomultipliers, where even tiny amount of chemically active molecules and ions created in an avalanche can destroy the photosensitive layer [86]. The use of GEM gaseous photomultipliers in pure $CF_4$ with reflective photocathodes, evaporated directly on the upper GEM electrode, opens new applications of windowless Cherenkov detectors, where both the radiator and the photosensor operate in the same gas [87]. As compared to CsI-multiwire detectors, the shielding power of the multi-GEM amplification structure considerably reduces photon (CsI is sensitive to $CF_4$ scintillation at 170 nm) and ion feedback at the photocathode, and therefore CsI aging in the gas avalanches. The compatibility of $CF_4$ with CsI-PC was demonstrated in one short aging test [88] (see Table 2); long-term aging studies with full size CsI-GEM detectors in $CF_4$, to ensure an efficient and reliable operation under LHC-like conditions, have to be foreseen.

## 5. Aging in Resistive Plate Chambers (RPC)

In the 1990's, RPC were proposed as an economical and proven technology ideally suited for precise time measurements (1 ns) on a large-area detector. Both the Belle and BaBar experiments have instrumented their flux returns with RPCs operated in streamer mode (see Table 3). One of the potential reason of RPC degradation is related to the fluorine compound - $C_2H_2F_4$, which is the main component of present RPC gas mixtures and can produce HF, especially with water vapour, under the electrical discharge [89]. Aging effects, presumably due to simultaneous etching of the glass by fluorine ions and deposition of avalanche products on the electrode surfaces, were first detected in Belle RPCs [50]. In dedicated laboratory tests, it was confirmed that after short operation of glass RPCs with ~ 1000 ppm of $H_2O$ in $C_2H_2F_4$-containing gas, partial/full recovery of efficiency is observed, while when glass RPCs operate in a wet gas for a long time the degradation becomes permanent. After the wet gas operation in $C_2H_2F_4$ both anode and cathode got damaged; the anode surface had a high level of fluoride, while the cathode lacked sodium. The authors explained the symptoms of aged chambers by spontaneous field emission of electrons from local deposits/tips on the cathode surface [90,91]. In one test, the area with reduced efficiency was clearly located along the preferential path of the gas flow [92]. It has to be noted, that ammonia flushing allows full recovery of the damaged glass chamber [91]. However, the glass RPC can suffer from the long-term instability due to the migration of alkali ions, which will lead to the permanent increase of the surface resistance. This phenomena prevented successful operation of ionic conductive D-263 glass and forced to choose electron conductive ordiamond coated glass for the MSGC substrate [3,8].

The initial problems observed in BaBar RPCs, made of Bakelite coated with linseed oil were related to the lack of polymerization of the linseed oil, electrochemical change of the linseed oil resistivity and formation of oil droplets under the influence of high temperature and high currents; more information can be found in [8,93,95]. Recent systematic R&D studies have shown that the bakelite volume and linseed oil surface resistance, which are the main parameters affecting the long-term performance of RPC, can be altered by the operation at increased temperature, total integrated current per unit surface, presence of water and complicated fluorocarbon chemistry. A steady increase of the total electrode resistance and therefore reduction of rate capability, as a result of the current flowing through the electrodes and especially due to the decrease of the water content in the bakelite under dry gas, was identified as the main aging effect in the oiled bakelite RPC [94]. Several studies indicated that the increase of surface resistivity can be greatly reduced by flushing the essential flow of humid gas through the chamber [94]. The ionic conduction mechanism of bakelite-based RPC, modulated by the presence of water was proposed by J. Va'vra [8]; further tests of the electrical conduction mechanism in the RPC might shed more light on the underlying physics processes and guarantee their reliable use under harsh experimental conditions. One of the main concern for the long-term RPC operation comes from the fact that the distribution of surface resistivity in large detector system is unpredictable as it depends on local variations of surface resistivity – which itself depends on consistency of the electrode surfaces as a function of the accumulated charge and level of humidity vs. gas flow. Another aging effect in RPC is related to the dramatical reduction of the linseed oil surface resistance due to the deposition of fluorine radicals formed from the $C_2H_2F_4$-based chemistry [8,89]; this effect may be significantly enhanced under the sparking even for avalanche mode operation. In addition, the smaller radiation damage in RPCs is expected for the reduced content of electronegative gas $SF_6$ in the mixture [94]. Another hypothesis to be investigated is that during operation with $C_2H_2F_4$ gas electrodes may accumulate photosensitive deposits; in this case UV light from avalanches will lead to the electron emission from the cathode, causing low detection efficiency and degrading chamber performance. Finally, future LHC experiments will operate RPCs in the avalanche mode, which is desirable in terms of total accumulated charge. However, much higher particle fluxes and appearance of discharges requires more study to find 'safe' operational parameters of the large RPC systems in the LHC environment. It should be stressed though, that the problems with RPCs are not 'classical aging effects', but rather unpredictable surface effects, related to the specific choice of construction materials and operating conditions.

Table 3. Summary of RPC operating conditions in the current and future HEP experiments.

| Experiment | Status | Electrodes material & resistivity | Gas mixture | Operation mode; charge/track | Particle rates ; Accumulated charge |
|---|---|---|---|---|---|
| L3 | Finished | Oiled bakelite $2*10^{11}$ Ωcm | $Ar/iC_4H_{10}/C_2H_2F_4$ (57:37:6) | Streamer | Consistent with cosmic rays |
| Belle | In progress | Float glass $10^{12}$ - $10^{13}$ Ωcm | $Ar/iC_4H_{10}/C_2H_2F_4$ (30:8:62) | Streamer | ~10-20 Hz/cm$^2$; |
| BaBar | In progress | Oiled bakelite $10^{11}$ - $10^{12}$ Ωcm | $Ar/iC_4H_{10}/C_2H_2F_4$ (60.6:4.7:34.7) | Streamer 1000pC/track | ~10-20 Hz/cm$^2$; <10 C/cm$^{-2}$ (in 2010) |
| ATLAS | Planned | Oiled bakelite $2*10^{10}$ Ωcm | $C_2H_2F_4/iC_4H_{10}/SF_6$ (96.7:3:0.3) | Avalanche 30 pC/track | <0.1 kHz/cm$^2$; <0.3 C/cm$^{-2}$ |
| CMS barrel | Planned | Oiled bakelite: $10^{10}$ Ωcm | $C_2H_2F_4/iC_4H_{10}/SF_6$ (96:3.5:0.5) | Avalanche 30 pC/track | <0.1 kHz/cm$^2$; <0.3 C/cm$^{-2}$ |
| ALICE | Planned | Oiled bakelite $3*10^9$ Ωcm | $Ar/iC_4H_{10}/C_2H_2F_4/SF_6$ (49:40:7:1) | Streamer | <0.1 kHz/cm$^2$; <0.2 C/cm$^{-2}$ |
| LHC-b | Abandon | Oiled bakelite $9*10^9$ Ωcm | $C_2H_2F_4/iC_4H_{10}/SF_6$ (95:4:1) | Avalanche 30 pC/track | 0.25-0.75 kHz/cm$^2$; 0.35-1.1 C/cm$^{-2}$ |

## 6. The Fundamental Limitations and Long-Term Performance of Gaseous Detectors at High Ionization Densities

The new generation of high-rate detectors of the LHC era has not only to cope with high dose rates, but also has to survive in the presence of highly ionizing particles, neutrons, low energy gammas and spallation products, with an average energy deposition 10-1000 times larger for MIP's. One of the most challenging applications of gaseous detectors, the efficient detection of minimum ionizing particles in a high intensity hadronic environment, might be limited in gain and rate by discharges. In wire-type detectors, the high counting rate lowers the gain due to the onset of space charge effects and this actually prevents sparking at high rates. Contrary to classical multi-wire chambers, in high-rate micropattern detectors the gain remains unchanged with rate. However, the maximum achievable gain before sparking drops with the counting rate and in the presence of heavily ionizing particles, an additional gain drop of 1-2 orders of magnitude may appear [96-98]. Owing to very small distance between anode and cathode for the fast removal of ions in the micro-pattern detectors the transition from proportional mode to streamer can be easily followed by the discharge, if the avalanche size (the product of ionization and gain) exceeds $10^7$-$10^8$ ion-electron pairs, the so-called Raether limit. Therefore, one should lower the gain at high rates for reliable operation. However, the ballistic deficit, i.e. the loss of the signal due to fast amplifiers and too short integration time in the LHC experiments at 25 ns beam crossings, does not allow to reach full efficiency to MIPs at sufficiently low gain. This situation may be aggravated in the $CF_4$-based mixtures, where due to the significant electron attachment and thus broader dynamic range of signals, a higher fraction of small amplitudes at the same gas gain is expected.

Therefore, if the detector operating point is set for efficient detection of relativistic particles, it is likely that a highly ionizing particle, releasing an exceptionally high amount of electron-ion pairs in the active gas volume, will produce a streamer, which may be followed by the onset of multiple streamers, sparks and discharges. Once the streamer establishes a conductive channel between anode and cathode the power dissipated in the spark is largely dependent on the effective discharge capacitance and the gas mixture. More robust detectors (GEM, Micromegas) are better suited for the harsh environments than MSGC systems. By an appropriate choice of the electrode configuration and the gas, these detectors can tolerate certain sparking rate, but avoid fundamental problems – big discharges causing permanent damages and subsequent detector failure.

After the many years of intensive research and development of radiation-hard gaseous detectors, an impressive variety of experimental data has been accumulated from laboratory tests and detectors installed at high energy physics facilities. However, there are many contradictory experiences obtained in seemingly identical conditions, which means we are still lacking even a macroscopic explanation of the behavior of irradiated gaseous detectors. It is now well established that – even if a low aging rate can be obtained in the laboratory with radioactive sources and otherwise very clean conditions – large-area detectors using the same mixture can fail due to severe aging after a relatively small beam exposure. Laboratory aging studies with radioactive sources suffer from one main limitation: they are obtained by irradiating a small part of the active surface of the chamber, and may be optimistic both in rate and ionization losses as compared to realistic beam operating conditions. Moreover, with a suitable selection of irradiated area of the detector, a wide range of performances can be obtained; this clearly demonstrates a danger to characterize the long-term performance of a large-scale detector by exposure to radiation of small areas only. Recent experimental results from hadron beam shows strong dependence of the detector lifetime upon basic macroscopic variables: ionization density, irradiation rate, particle type and energy, high voltage (gas gain), size of the irradiated area and gas flow [5]. These data emphasize the importance to study the radiation hardness of the 'large-scale' prototype detector under conditions as close as possible to the real environment.

### *6.1. Aging vs. ionization density, irradiation rate and particle type*

The experimental dependence of detector lifetime on ionization density, and in particular on irradiation rate, particle type and energy, can be related to the charge density and total dissipated energy in the detector from the incoming particles. It is well known that for many gaseous detectors (single wire, MWPC), the counting rate capabilities are limited by space-charge effects in the avalanches, which can either lead to gain reduction or, for higher applied voltages (when the total charge exceeds Raether limit) to the formation self-quenching streamers. Smaller gain losses for larger current densities, observed in laboratory tests with radioactive sources or X-rays [1-4], were attributed to the onset of space charge effects, which reduce the electron energy in the avalanche, thus decreasing the density of ions and radicals in the avalanche plasma (see Fig.3). However, the transition between proportional and streamer mode in all types of gaseous detectors, as well as the streamer pulse charge itself, depends not only on the high voltage but also on the primary ionization density [100,101]. Due to the high density of primary electron-ion pairs produced by α's, the Raether threshold for the formation of streamers along the α-track may be reached even with moderate multiplication factors; the streamer produces a densely ionized and therefore low resistivity plasma medium between anode and cathode, which in some cases may lead to spark events. Each spark is a potential danger from the aging point of view because it can 'prime' the electrode surface for subsequent buildup of deposits. Any tips left from sparks on electrodes create local enhancement of electric field at the edges of imperfections. The self-sustained Malter current and sporadic electron jets from microscopic dielectric insertions on the cathodes [102,103] may be as significant as heavily ionizing particles, since the ions of the second generation proceed directly back to the same cathode spot along the same electric field lines that the secondary electrons followed, thus creating feedback loops and facilitating the discharge propagation along these low resistivity channels. The number of molecules and the extent to which they 'crack' along these densely ionized conductive plasma chains can be greatly increased compared to ordinary wire avalanches, thus significantly reducing detector lifetime. Additionally, at large current densities (>100 nA/cm) and in presence of highly ionizing background, the streamers start overlapping in space and time effectively adding their respective charges; this effect is proportional to the irradiated area and may contribute to a greatly enhanced polymerization rate in large systems (see Fig.3). According to this interpretation there could be some limiting current density which each detector can accept before the appearance of electrical breakdown [97].

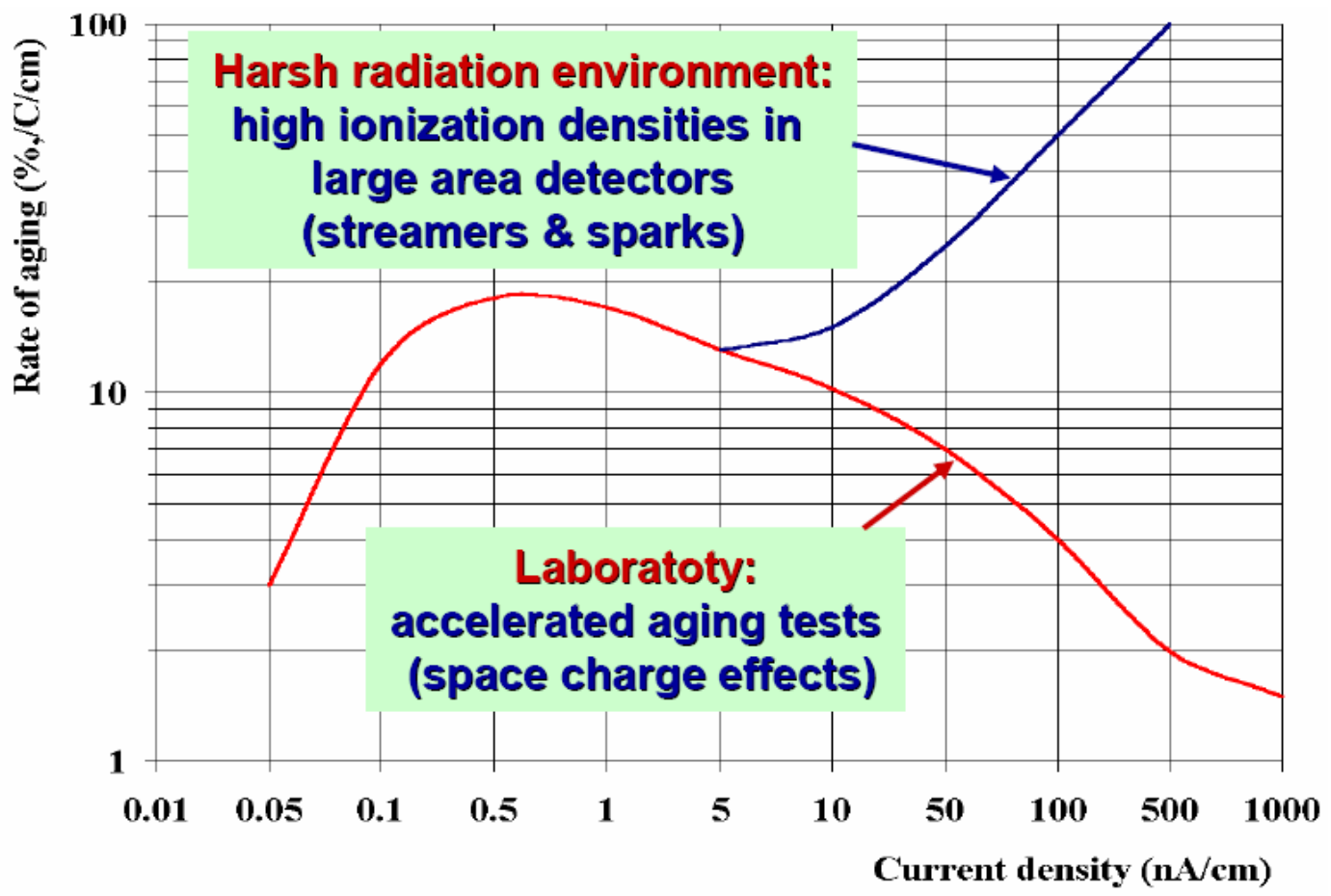

Fig.3. A phenomenological model of the aging rate dependence from the current density in the laboratory tests (with radioactive sources or X-rays) and under harsh radiation LHC-like environment [104].

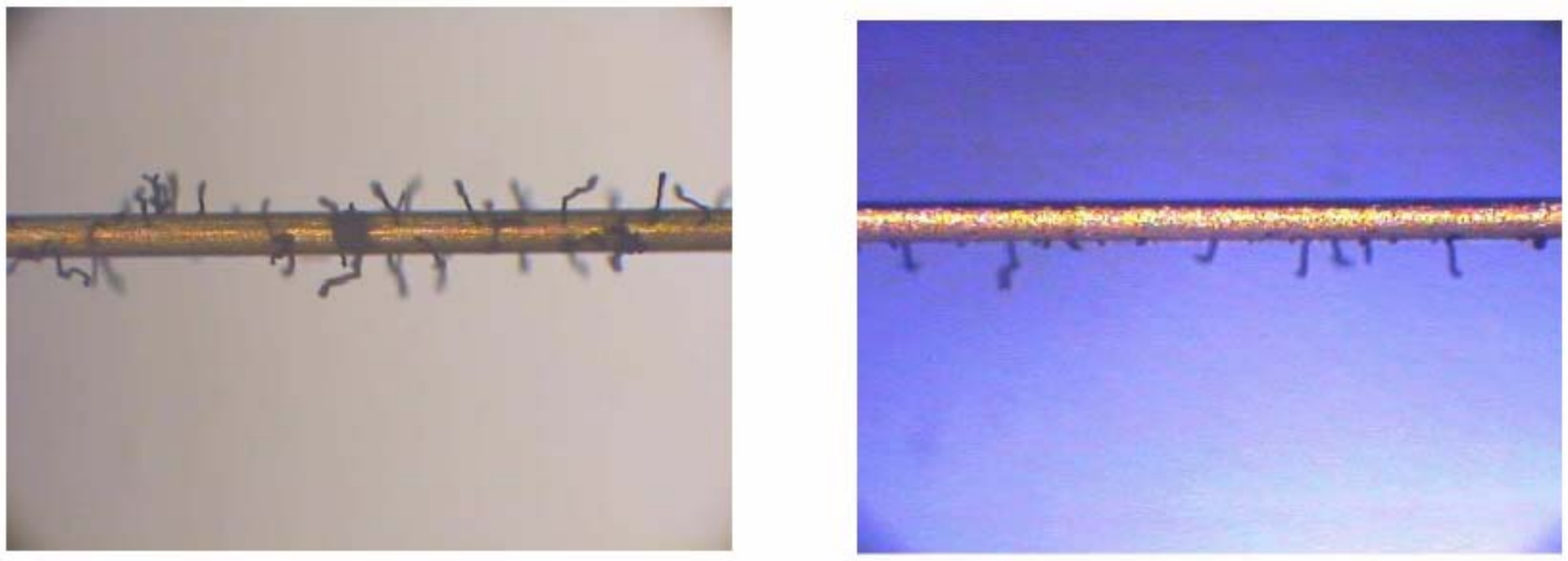

Figure 4. (Left) Deposits on the anode wire after exposure to α's in Ar/$CF_4$/$CH_4$(74:20:6). (Right) Whiskers on the wire irradiated over half of its cell width with α's in Ar/$CF_4$/$CH_4$(74:20:6) [35].

With energy losses per α-track that can be ~$10^3$ times larger than for primary ionization with low density (X-rays or MIP), aging shows up as hair-like deposits randomly distributed within the irradiated area (see Fig. 4). The aging rate in Ar/$CF_4$/$CH_4$ (74:20:6) was found to be two orders of magnitude higher in a 100

MeV α-beam than in a small-scale laboratory test with a $Fe^{55}$ source; both experiments were performed at similar current densities ~ 700 nA/cm [35]. Of special interest were the wires irradiated over half of their cell width; the 'whisker'-type deposits were built only on one side (along the α-particle ionization tracks) towards which all primary electrons drifted (see Fig.4). These results clearly indicate that the aging properties can not be predicted solely on the basis of atomic composition ratios in the mixture, without taking into account the actual operating conditions. Since discharges, sparks and electrical breakdowns could partly simulate environments with large radiation doses, it makes sense to perform special tests irradiating large area prototype chambers under high intensity of heavily ionizing particles to eliminate gases and construction materials which may promote the growth of polymers and thus examine the 'safe parameters' at which detector can be operated. However, these conditions are more severe than environment expected at LHC, where pions are dominating at small radii, being accompanied by heavily ionizing particles. It has to be noted that the determination of the accumulated charge, by integrating the continuously recorded anode current divided by the entire irradiated surface ('current method'), can be only used if the spark and discharge rate are negligible; otherwise the current is dominated by huge charge doses at local 'spark points' on electrodes (e.g. tips, polymer deposits, dust particles, fingerprints). The 'current method' may also severely underestimate the collected charge at local chamber points after the appearance of the positive feedback and self-sustained Malter discharge at such distinct spots. It has to be emphasized, that no conclusions about the detector longevity in high intensity hadron beams can be drawn from the small-scale laboratory tests performed with radioactive sources or X-rays at huge accumulation doses >1 μA/cm (>100 nA/$mm^2$ for MPGD); the final prove of the radiation hardness of gas detector can be only acquired with exposures in realistic conditions (both in particle composition and ionization density).

### *6.2. Aging vs. high voltage/gas gain/electric field strength*

The experimental dependence of aging properties on the high voltage may be related to the dependence from the electric field strength on the anode surface, which enhances inelastic processes (excitation, ionization) and determines the electron temperature in the avalanches. A lower amplification field also strongly reduces the UV light emission in the avalanches and thus the strength of the various secondary phenomena on the nearby electrodes [99,105]. Fig.5 shows an example of nearly exponential decrease of detector lifetime with increase of anode voltage [56]. The aging rate may be also accelerated by avalanche formation at the cathode surface, if the electric field there exceeds ~10-20 kV/cm. Moreover, the lower the electric field at the cathode surface the more accumulated charge is required to deposit an insulating film at the cathode, since most of polymerizable species are produced in the high field region close to the anode and delivered to the cathode by electrostatic forces (due to positive charge or dipole moment) [1].

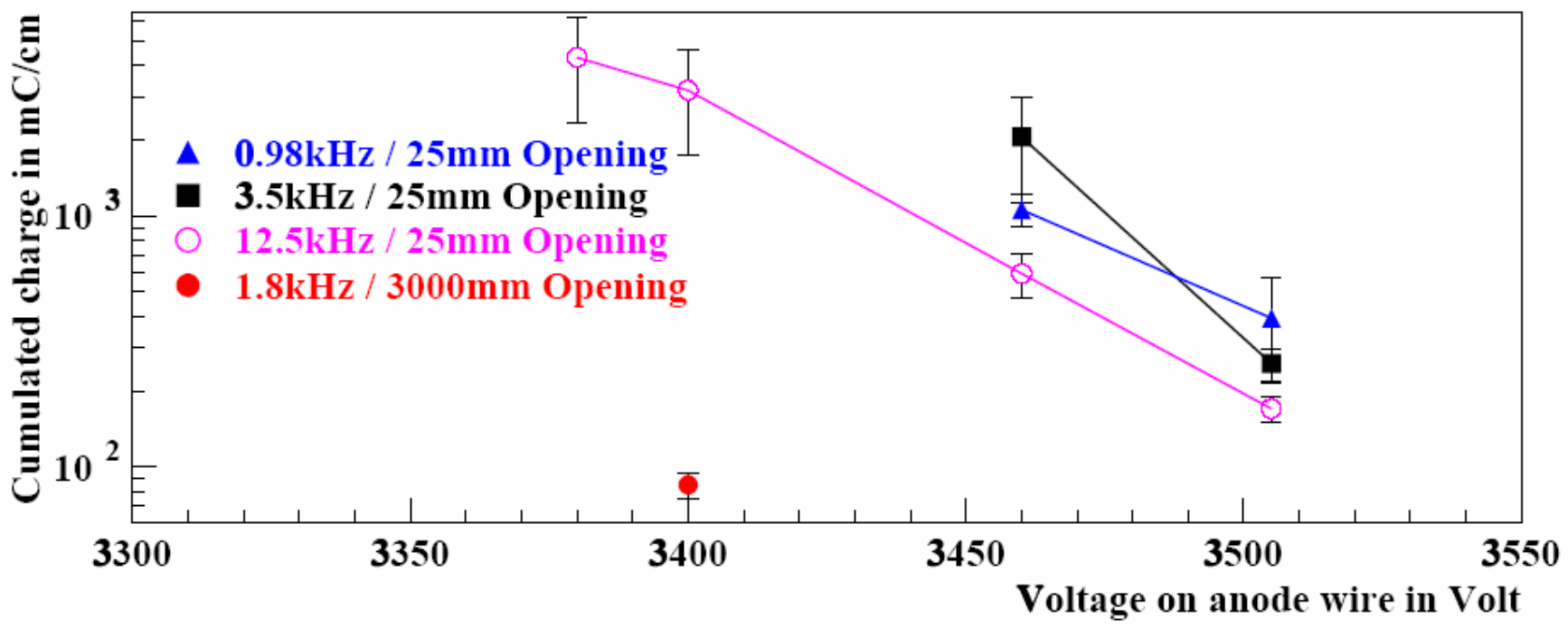

Figure 5. Dependence of lifetime of ATLAS muon drift tubes irradiated in $Ar/CH_4/N_2/CO_2$ (94:3:2:1) as a function of high voltage for different particle fluxes [56].

In recently developed GEM devices, a reduction in the electric field strength (voltage drop and thus a lower amplification factor across a single GEM) was found to be a major factor which increases the reliability and radiation hardness of a triple-GEM detector, compared to double GEM, without sacrificing the total gain of the system [68,70]. Superior aging properties of GEMs, which are rather insensitive to aging compared to wire-type detectors, may be explained by the separation of multiplication and readout stages (the gas amplification is localized inside GEM holes) and by the smaller rate of polymerization of impurities in the much lower electric field in this region (50-100 kV/cm), compared to the anode wire surface field ~250 kV/cm [70]. The lower electric field strength and larger avalanche charge spread over the whole multiplication region in the parallel plate mode (50 kV/cm), instead of just around a thin anode wire, may also lead to the high radiation hardness of Micromegas.

Finally, the appearance of discharges depends strongly on the detailed electrical field configuration in the detector, on any imperfections in the production process and on the composition of the radiation field (e.g. particle type, flux and energy). In addition, the gain-voltage characteristics is steeper in poorly quenched gases (e.g. $Ar/CO_2$), revealing the essential role of manufacturing quality required for a uniform performance of large size detectors [98]. However, what is even more important is the topological layout under irradiation. If the conductivity of gas, due to the radiation-induced breakdown of gas rigidity and onset of large currents, becomes higher than the one of the insulator at the 'triple junctions' (singular points, where metal electrode structures touch the insulator in the counting gas) the enormous field enhancement close to dielectric surface is capable of inducing electrical breakdown during the avalanche multiplication process [106]. Special attention should be given to the detector rim, the region most prone to discharges. Therefore, the electric field strength in the detector at high ionization densities is determined not only by capacity matrix but also by currents [106]. Since the radiation damage may be accelerated by sparks, discharges and photoelectric feedback, in designing of gas detector one should aim at the lowest possible electric field distribution (gas gain) compatible with the functional requirements.

### *6.3. Aging vs. size of the irradiated area and gas flow*

The dependence of the detector lifetime on the size of the irradiated area was observed in all types of gaseous detectors (e.g. wire counters, MPGD, RPC). The increase of the aging rate in the direction of the serial gas flow, as shown in Fig.6, means that aging should be viewed as a non-local and intensity dependent phenomena. Some of the long-lived aggressive free radicals may diffuse within the irradiated detector area and react with other avalanche generated dissociative products, construction and electrode materials thus enhancing aging effects with increasing usage of the gas. These observations seem to be the most critical when trying to extrapolate the aging behavior from laboratory tests to large-scale detectors and require to avoid gas distribution systems that supply many chambers by a serial flow.

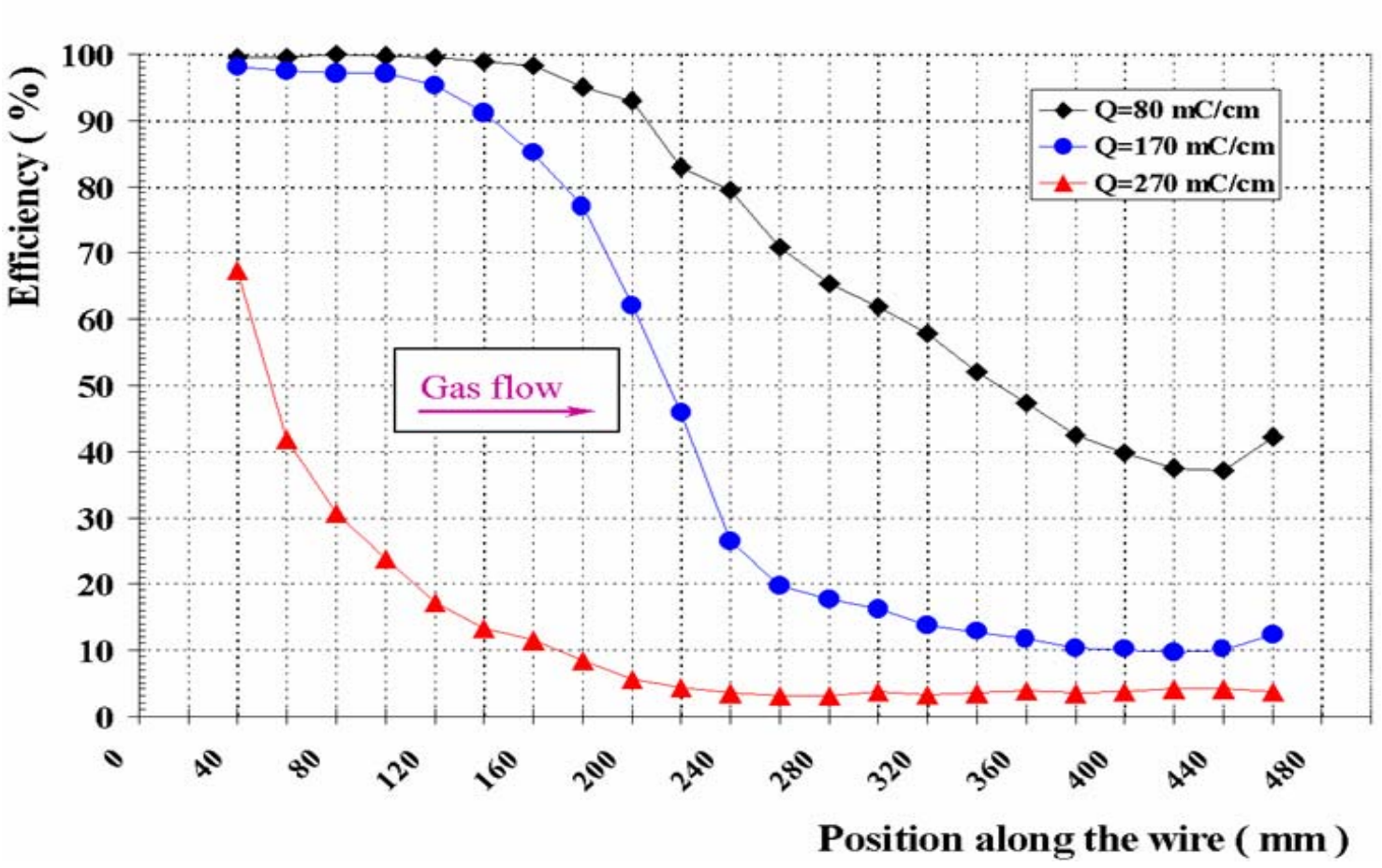

Figure 6. Progressive deterioration of efficiency to MIP particles in the direction of the serial gas flow in the HERA-B prototype muon chambers for three different accumulated charges: 80 mC/cm, 170 mC/cm and 270 mC/cm of the wire in an $Ar/CF_4/CH_4$ (67:30:3). The whole wire length (500 mm) was irradiated. The maximum of irradiation intensity corresponds to x~250 mm. [35].

It is rather intuitive that the aging rate typically decreases with increasing gas flow, since it tends to sweep away the harmful radicals and ions, leading to polymer growth [3,5]. However, this trend is not always correct, for example, if the polymerization is triggered by the impurities in the input flow stream. By increasing the gas flow, more of this ingredient is available to speed up the aging process; such tendency was in some cases observed in gas detectors [16].

### *6.4. Some Useful Guidelines for the Construction and Operation of Gaseous Detectors in High Luminosity Experiments*

Over the last decade considerable progress has been made on understanding of basic rules for construction and operation of radiation-hard gaseous detectors, which might help to prevent or at least to suppress the aging rate to an acceptable level [1-3,5]; the most important ones are summarized as follows.

1) Carefully choose construction materials - radiation hardness and outgassing properties are of a primary importance. There are clearly many ‘bad’ and a lot of ‘usable’ materials - perform tests to match your specific requirements.

2) Only a limited choice of aging resistant gases can be successfully used in the high intensity experiments: noble gases, $CF_4$, $CO_2$, $O_2$, $H_2O$. Hydrocarbons are not trustable for long-term high rate experiments. Operational problems could be aggravated by using poor quencher $CO_2$ and by the very high aggressiveness of $CF_4$ dissociative products.

3) Validate your assembly procedures, which must include maximal cleanliness for all processes, quality control for all system parts and personnel training (no greasy fingers, no polluted tools, no spontaneously chosen materials installed in the detector or gas system in the last moment, before the start of real operation).

4) Carefully control any anomalous activity in the detector, such as dark currents, changes of anode current and remnant activity in the chamber when beam goes away. If aging effects are observed despite all precautions, add oxygen-based molecules to inhibit/relief/cure polymerization of hydrocarbons; operation with $CF_4$ decreases the risk of Si polymerization.

5) Special care has to be taken to the surface conductivity of electrodes, since it is closely related to the capability of operating at large ionization densities. The resistivity of the microscopic insulating layer on the metallic cathode surface defines the maximum rate capability of the detector before the onset of electron field emission from the cathode, which starts if the rate of ionic charge neutralization across the dielectric film is smaller than the rate of ion charge built-up. Detectors, which use insulators (MSGC, CsI, RPC) may face a new domain of aging: radiation-induced increase of surface resistivity of electrodes and supporting structures due to ionic currents; this may trigger Malter-type breakdowns [8].

## 7. Conclusions and Perspectives

A centenary after the invention of the basic principle of gas multiplication, gaseous detectors are still the first choice whenever the large area particle detection of medium spatial accuracy is required. The recent invention of the micro-pattern gas detectors promises to extend the applicability of gaseous detectors to the precision tracking at high counting rates in hostile environments, an area previously accessible only to silicon detectors. The main properties of gaseous and silicon microstrip detectors, fulfilling all necessary requirements concerning high-rate capability and radiation hardness in the intermediate region from 20 cm up to 2 m from the beam pipe, are summarized in Table 3. Both Micromegas and GEM detectors already achieve a position resolution of better than 70 μm, timing resolution of ~10 ns and signal to noise ratio (S/N) of more than 20 at full efficiency in high-intensity beams. In the laboratory tests using narrow readout strips and low-diffusion gas fillings, a time resolution ~ 5 ns and the spatial resolution of ~ 15-30 μm, close to that of the silicon detectors, was measured. The most exciting prospect for future developments could be pixel readout structures for these kinds of devices.

Table 4. Basic parameters of the tracking detectors. For MPGD the spatial ($\sigma_x$ ) and time resolution ($\sigma_t$) are given for laboratoty (high intensity beam) measurements. The $X_0$ refers to the amount of material in one tracking station (two layers of straws are considered in one tracking station). (a) The rate capability of Si μ-strips is limited by properties of readout electronics.

| Detector | Radiation Hardness | $\sigma_x$, μm | $\sigma_t$, ns | S/N | $X_0$, % | Rate capability Hz/mm$^2$ | Size of detector |
|---|---|---|---|---|---|---|---|
| **Straw tubes** | **> 10 C/cm (>$10^{12}$ MIPs/mm$^2$)** | **100** | **50** | **15-20** | **0.2** | **~ $10^4$** | **300*0.4 cm$^2$** |
| **GEM** | **> 20 mC/mm$^2$; (> 6*$10^{11}$ MIPs/mm$^2$)** | **30(70)** | **5(12)** | **>20** | **0.4** | **~5*$10^5$** | **30*30 cm$^2$** |
| **Micromegas** | **20 mC/mm$^2$ (6*$10^{11}$ MIPs/mm$^2$)** | **15(70)** | **5(10)** | **>20** | **0.4** | **~5*$10^5$** | **40*40 cm$^2$** |
| **Si μ-strips (300 μm)** | **3*$10^{12}$ (24 GeV protons)/mm$^2$** | **< 10** | **~20** | **15-20** | **1.2** | **(a)** | **8*8 cm$^2$** |

The dramatic increase in charge, which is expected to be accumulated on sensing electrodes in the new high-rate experiments, posed much more stringent constraints on the radiation hardness of materials and gas mixtures, and basic rules for construction and operation of gaseous detectors, than even thought before. No general recipe can be given for the radiation hardening of the detectors. The aging tests should include an extended study of 'large-scale' final prototype chambers, exposed over the full area to a realistic radiation profile (particle type and energy, ionization density, irradiation rate). It is of primary importance to vary the operating parameters systematically in order to investigate their possible influence on the aging performance. In order to exclude statistical fluctuations of unknown nature and to provide the reliable estimate for the detector lifetime, the radiation tests should be carried out with several detectors irradiated under identical conditions. Ten years of intensive research aiming to match the needs of high luminosity colliders have demonstrated that if properly designed and constructed, gaseous detectors can be robust and stable in presence of high rates and heavily ionizing particles.

## Acknowledgments

I would like to thank Eugenio Nappi and Jacque Seguinot for the invitation to the Erice workshop and for their great hospitality. I am especially grateful to Fabio Sauli (CERN) for many helpful suggestions, stimulating discussions and careful reading of this manuscript. I would like to thank to Ulrich Parzefall (Freiburg University) and Mar Capeans (CERN) for reading and correcting this manuscript.